\documentclass[article,amssymb,showkeys]{revtex4}
\usepackage{graphicx}
\DeclareGraphicsExtensions{.eps}
\usepackage{dcolumn}
\usepackage{bm}

\newcommand{\To}{\longrightarrow}

\begin{document}

\title{The power law character of off-site power failures}
\author{A. John Arul \footnote{Corresponding author's e-mail:
arul@igcar.ernet.in}, C. Senthil Kumar$^\natural$, S. Marimuthu
and Om Pal Singh} \affiliation{\it{Reactor Physics Division,
Indira Gandhi Centre for Atomic Research,$^\natural$AERB-Safety
Research Institute
\\ Kalpakkam, 603 102, India}}
\date{\today}

\vspace{6mm}

\begin{abstract}
\noindent A study on the behavior of off-site AC power failure
recovery times at three nuclear plant sites is presented. It is
shown, that power law is appropriate for the representation of
failure frequency-duration correlation function of off-site power
failure events, based on simple assumptions about component
failure and repair rates. It is also found that the annual maxima
of power failure duration follow Frechet distribution, which is a
type II asymptotic distribution, strengthening our assumption of
power law for the parent distribution. The extreme value
distributions obtained are used for extrapolation beyond the
region of observation.
\end{abstract}


\maketitle

\noindent \textbf{1. Introduction} \\
\indent Estimation of off-site power failure characteristics is
important for the safe design and operation of nuclear power
plants. The emergency power supply requirements are based on the
ability to forecast the maximum credible loss of off-site power
(LOSP) duration during the life time of the plant. The observed
data is available only over a period of less than two decades and
scarce on long duration failures. We address two questions: first,
what is the nature of the parent distribution, i.e., the form of
frequency-duration correlation function and second how to
extrapolate for failure durations longer than the observed
maximum.

First we show, based on simple assumptions about the off-site
power supply system component failure and repair rates, that the
frequency-duration correlation function is a power law of the
form, $y\propto x^{-k}$. Next, it is established that the
distribution of annual maxima, of observed failures follow Type II
asymptotic distribution (Frechet distribution, see Johnson 1970).
This result strengthens the premise, that the parent distribution
for power failure duration-correlation is power law in character.
Extreme value analysis is also used to make extrapolations for
more number of years than observed, for instance to estimate the
most probable loss of off-site power duration in say, 50 or 100
years.

\vspace{5mm}
\noindent \textbf{2. Off-site power failure frequency-duration function}\\
\indent The off-site AC power at a plant may be lost due to supply
failure from the power grid or because of plant centered equipment
failure like station transformer, feeders, breakers etc. The grid
failure could be a minor grid disturbance or grid collapse. The
grid power may also fail due to severe environmental conditions.
Baranowsky, (1988) has used Weibull functions of the form $(
\lambda_i e^{(-\alpha_i t^{\beta})} )$ for representing the
frequency of off-site power failure events of each type, exceeding
a given duration. Although the fits are good, it would be better
if some physical basis could be established for the fitted
function. Further considering the scarcity of data, instead of a
series of stretched exponentials a single function would be
preferable. It is postulated here that there is a mixture of power
failure rates $\lambda_i $ with a certain probability density
$g(\lambda_i)$ as $\lambda_i \in \{\lambda_i | i=1..N\} $.
Similarly for the repair rates $\mu_i \in \{\mu_i | i= 1..N\} $.
That is, there are different kinds of equipment each with a
specific $\lambda_i$ and $\mu_i$. Then the probability of
observing a failure of duration $ t' \ge t $ is

\begin{equation}
P( t' \ge t) = \sum_{i} { g(\lambda_i) \exp(-\mu_i t) }
d\label{eq:sum}
\end{equation}

where $e^{(-\mu_i t)}$ is the probability that the failed
component is not restored in time t, assuming constant repair rate
$\mu_i$. Since more frequent failures are generally repairable in
shorter times compared to rare failures which take more time to
set right, it is assumed that $ \mu \propto \lambda $, with this
assumption in equation \ref{eq:sum}, and going over to the
continuous limit we get,
\begin{equation}
P( t' \ge t) = \int { g(\lambda) e^{(- c \lambda t)} d\lambda}
\label{eq:integ}
\end{equation}
Where $c > 0$ is a constant of proportionality. The form of $g$ is
not known, however it is known (Johnson, 1970) that if gamma
distribution $$ g(\lambda) = \alpha e^{(-\alpha \lambda)} (\alpha
\lambda)^{(k-1)}/\Gamma(k)$$ for constants $\alpha,k > 0 $, is
assumed for $g$, then Pareto distribution is obtained as the
cumulative distribution function $F(t) = 1- P(t'\ge t)$. That is
the solution of equation \ref{eq:integ} is
\begin{equation}
P( t'\ge t) =  {1 \over(1+{c\over \alpha} t)^k}
\end{equation}

For large $t$, $P(t)$ goes like $P(t) \backsim(\theta/t)^k; k > 0
;\theta > 0; t\geq0 $, where $\theta = \alpha/c $ and
\begin{equation}
 F(t)\approx 1-(\theta/t)^k   \label{eq:pl}
\end{equation}
This family includes Cauchy, $t$ and $F$ distributions. The
function P(t) is plotted for data observed at two plant sites,
over a period of 15-20 years, in Fig. 1 and 2. The plots show P(t)
versus LOSP time $t$ in Log-Log scale. The linear fits are shown
alongside, which confirm the power law character of the
distribution. In Fig. 1 there is considerable deviation from
straight line for the first few points. This is due to the fact
that for small values of $t$, i.e., for ${c\over \alpha} t << 1$,
P(t) is virtually constant, and the power law dependence manifests
fully for large $t$. The LOSP frequency-duration correlation
function has the same time dependence as it will be different from
$P(t)$ only by a constant factor.

\vspace{5mm}
\noindent \textbf{3. Extreme value analysis of LOSP duration}\\
\indent The distribution of maximum of $m$ observations of the
random variable $t$, distributed as F(t) is,
$$
\lim_{m\To\infty} F(t)^m = G(t)
$$
The non-trivial limit distribution G(t) is obtained by appropriate
scaling of the variable $t$,
$$
\lim_{m\To\infty} F((t-b_m)/\theta_m)^m = G(t)
$$
when $F(t)$ is of the form of equation \ref{eq:pl}, the asymptotic
limit, with $b_m = 0$ is
\begin{equation}
G(t)=e^{-(\theta/t)^k} \label{eq:frechet}
\end{equation}
which is one of the three possible asymptotic limits, that is, a
type II extreme value distribution (Fisher et al., 1928 and
Gumbel, 1954) also known as Frechet distribution. For finite $m$
equation \ref{eq:frechet} is written as
$G_m(t)=e^{-(\theta_m/t)^k}$.

Extreme value analysis of LOSP data collected in (Theivarajan
1999, Marimuthu et al., 2000 and Kumar et al., 2002), is done as
follows. The annual maximum $t_i$ for each location is arranged in
ascending order as, $t_1 < t_2 < ...< t_N$. The plotting points
F(i) are calculated as
$$ F(i) =(i-0.3)/(N+0.4) $$(See Karl Bury, 1999), where N is the
number of years over which the data is collected. The $F(i) $
approximate the median of the distribution free estimate of the
cumulative distribution function. When $-ln(-ln(F(i))) $ are
plotted against $ln (t)$,  where $t$ is the power failure
duration, straight lines are obtained as depicted in Figs. 3, 4
and 5, as expected from equation \ref{eq:frechet}. The exponents
$k$ obtained from the slope, is shown in the respective figures.

The distribution of maximum in any $n$ years will be
\begin{equation}
 G_n(t) = G_m(t)^n = e^{-(\theta_n/t)^k}
\label{eq:extrm}
\end{equation}
where $$\theta_n = \theta_m n^{(1/k)}$$ \\
The most probable maximum in $n$ years of observation, obtained
from equation \ref{eq:extrm} is,
$$
\widehat{t} = \theta ( {n k\over{k+1}})^{(1/k)} \label{eq:tmp}
$$

The most probable maximum $\widehat{t}$ is plotted in Fig. 6, for
the three locations. The relatively higher value of most probable
maximum for TAPS and FBTR cases is due to higher incidence of
recorded on-site power distribution equipment failure. From Fig. 6
we can infer that, for instance a 20 h LOSP extrema is likely in
TAPS site in 15 y, whereas events of such duration are expected to
occur only in 50 years and 100 years respectively for the FBTR and
MAPS sites. These conclusions are based on the data collected from
these sites and results are indicative only.

\vspace{5mm}
\noindent \textbf{4. Conclusion} \\
\indent The nature of the relationship of off-site power failure
duration and its frequency is studied and it is found to have
power law dependence. Plausible physical basis for the observed
behavior is given. The power law nature of asymptotic behavior is
confirmed by performing an extreme value analysis of the data. The
extreme value distribution is also used to extrapolate beyond
observed LOSP durations.

It is interesting to note that, power laws have been observed in a
variety of natural and man made settings like, frequency
distribution of words (Zipf, 1949), distribution of incomes,
earthquake magnitudes and recently in internet topology (Faloutsos
et al., 1999), to name a few. Based on the intimate connection
between power laws and the ideas of self organized criticality
(Bak et al., 1987), it can be said that off-site power
distribution system seems to evolve into a critical state, where
failures of longer duration are not infrequent, as one would
expect for exponential distribution. Although it is convenient not
to consider cutoffs to failure duration from a theoretical
perspective, it would be better if methods could be devised to
include cutoffs considering the finiteness of the system.


\vspace{10mm}

\noindent \textbf{References}

\begin{enumerate}
\item Bak P., Tang C., and Wiesenfeld K. 1988. Self-Organized
Criticality, Phys. Rev. A, 38, 364.
\item Baranowski, P. W. 1988. Evaluation of Station Blackout
Accidents at Nuclear Power Plants, NUREG-1032, USNRC, Washington.
\item Faloutsos M.,  Faloutsos P., and Faloutsos C. 1999. On Power-Law
Relationships of the Internet Topology, SIGCOMM '99, pp. 251-262.
\item Fisher R. A. and Tippet L.H.C., 1928, Limiting forms of the
frequency distributions of the largest or smallest members of a
sample, Proceedings of the Cambridge Philosophical Society 24, pp.
180-190.
\item Gumbel, E. J. 1954. Statistical Theory of Extreme
Values and Some Practical Applications, National Bureau of
Standards Applied Mathematics Series 33, Washington.
\item  Johnson N. L. and Kotz S. 1970. Distributions in
statistics continuous univariate distributions-1, John Wiley \&
Sons, New York.
\item Karl Bury, 1999. Statistical Distributions in Engineering,
Cambridge University Press.
\item Kumar, C. S., Arul, A. J., Anandapadmanaban, B., Marimuthu, S. 2002. Estimation of Station Blackout Frequency in
FBTR, ROMG/FBTR/S-AX-01/52000/SAR-38.
\item Marimuthu, S., Theivarajan, N., Kumar, C. S., and John Arul, A.
2000. Statistics of Loss of Off-Site Power at Kalpakkam, Rev. A.,
PFBR/01160/DN/1000, Kalpakkam, India.
\item Theivarajan, N. 1999. Recommended Design Basis Data for the Loss of
Off-site Power, IGCAR Internal Report, PFBR/51100/DN/1001,
Kalpakkam, India.
\item  Zipf, G. K. 1949. Human Behavior and Principle of Least Effort:
An Introduction to Human Ecology, Addison Wesley, Cambridge,
Massachusetts.

\end{enumerate}
\clearpage

\DeclareGraphicsRule{.ps}{eps}{}{}

\begin{figure}
\includegraphics[width=3in,height=2in]{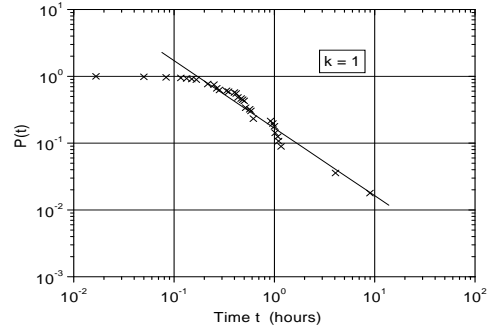}
\caption {\label{a} P(t) versus loss of off-site power duration
$t$ for site Kalpakkam-MAPS}
\end{figure}
\begin{figure}
\includegraphics[width=3in,height=2in]{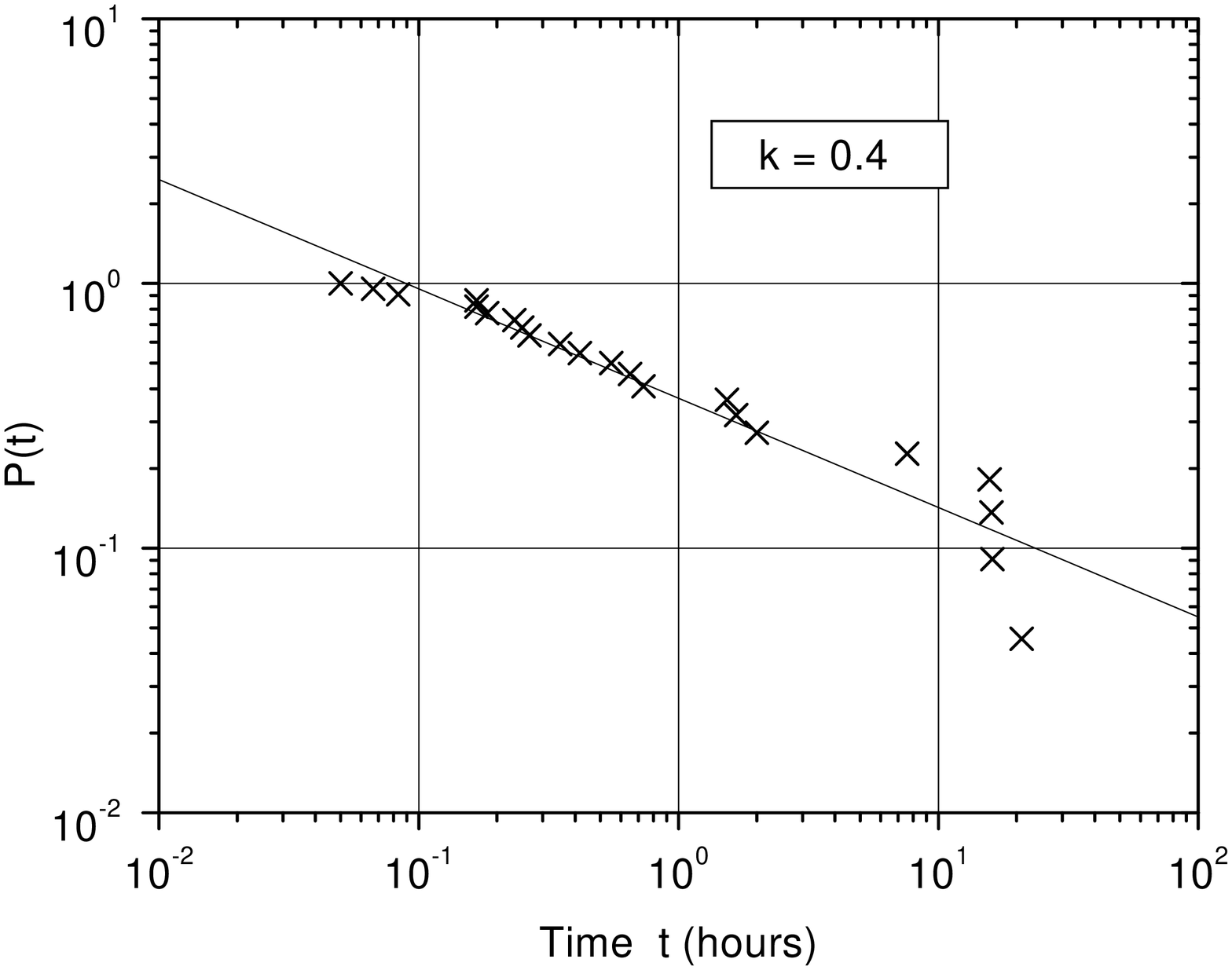}
\caption {\label{b} P(t) versus loss of off-site power duration
$t$ for site Kalpakkam-FBTR}
\end{figure}
\begin{figure}
\includegraphics[width=3in,height=2in]{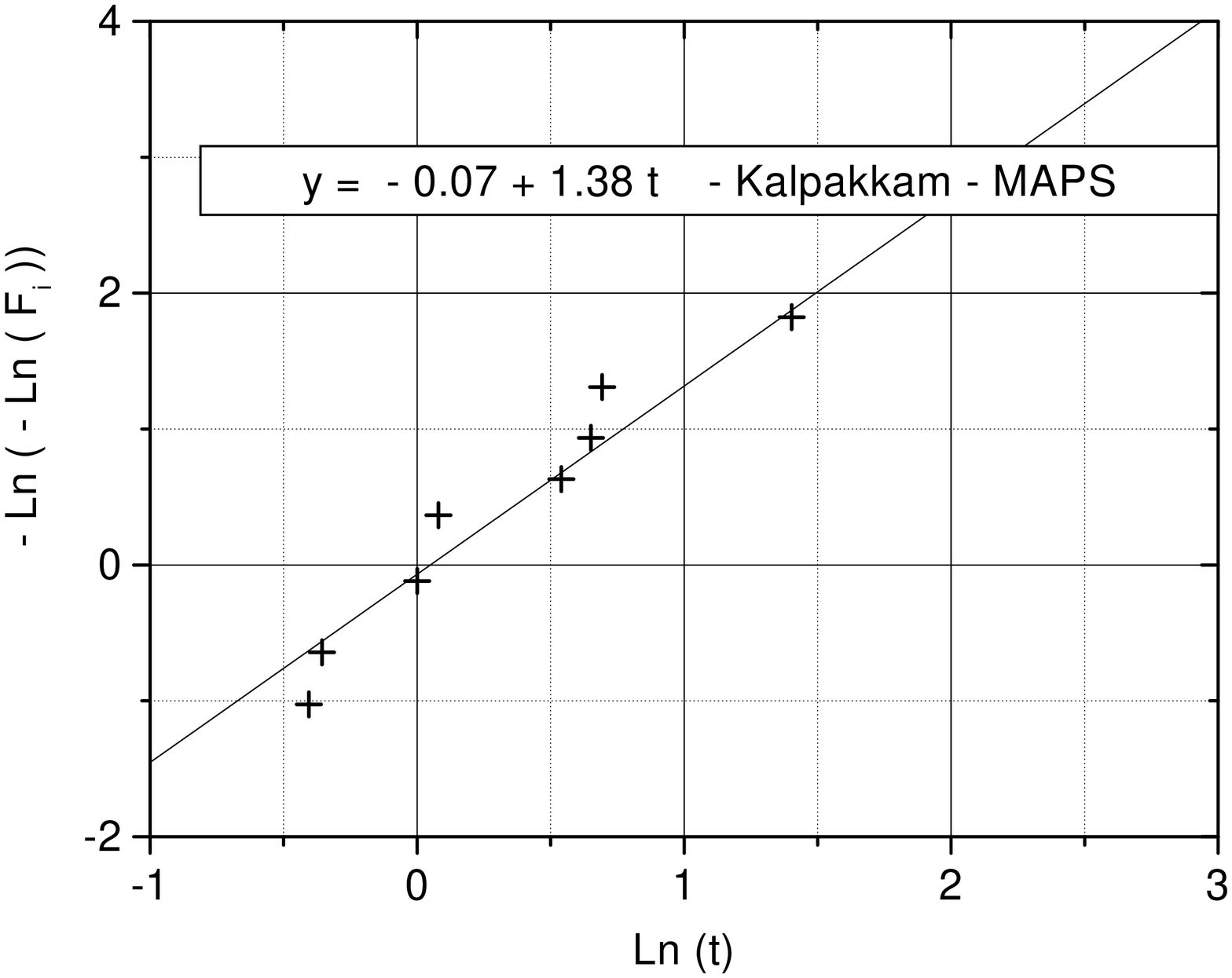}
\caption {\label{c} Log Log of $F_i$ versus Log of maximum annual
loss of off-site power duration for site Kalpakkam-MAPS}
\end{figure}
\begin{figure}
\includegraphics[width=3in,height=2in]{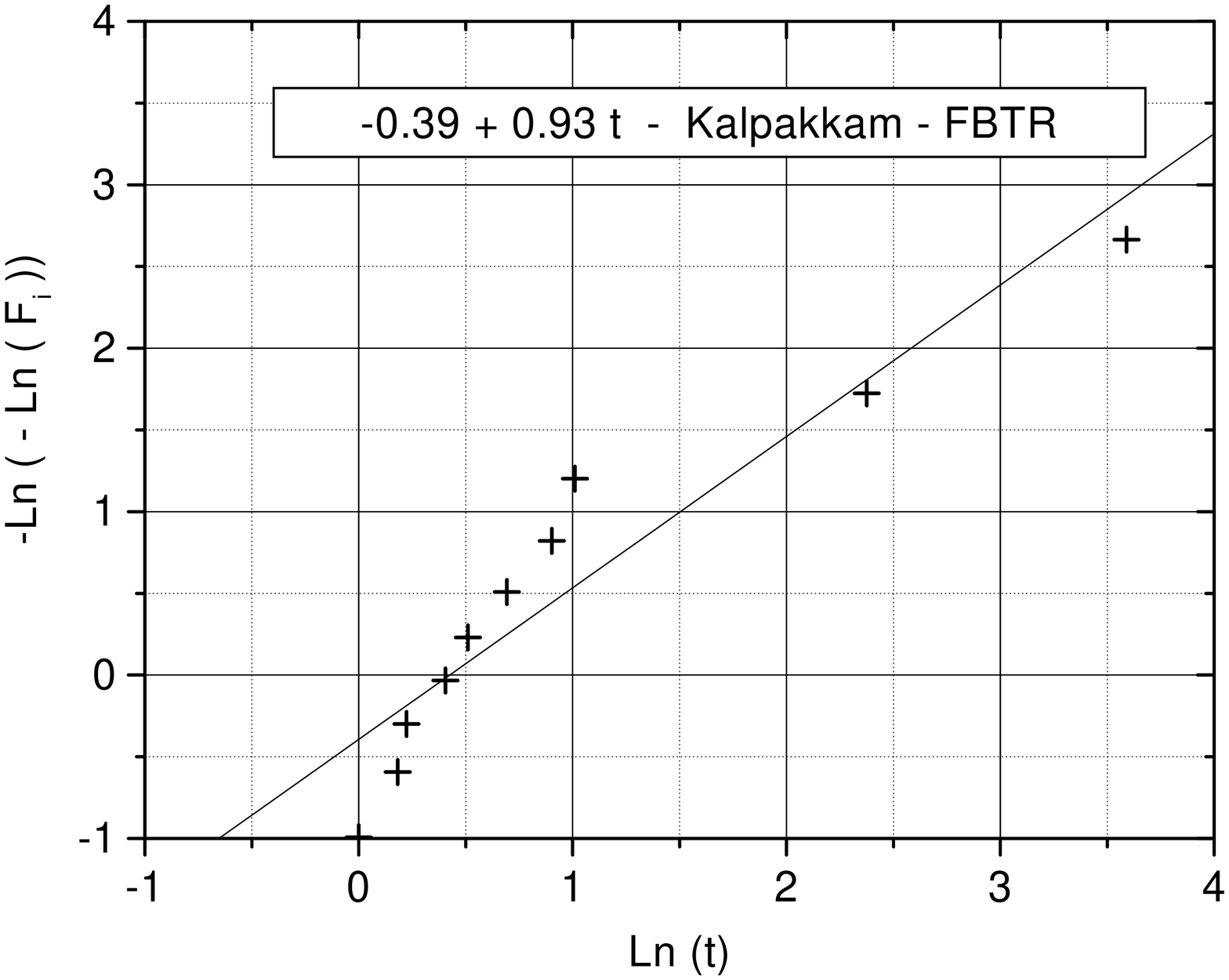}
\caption {\label{d} Log Log of $F_i$ versus Log of maximum annual
loss of off-site power duration for site Kalpakkam-FBTR}
\end{figure}
\begin{figure}
\includegraphics[width=3in,height=2in]{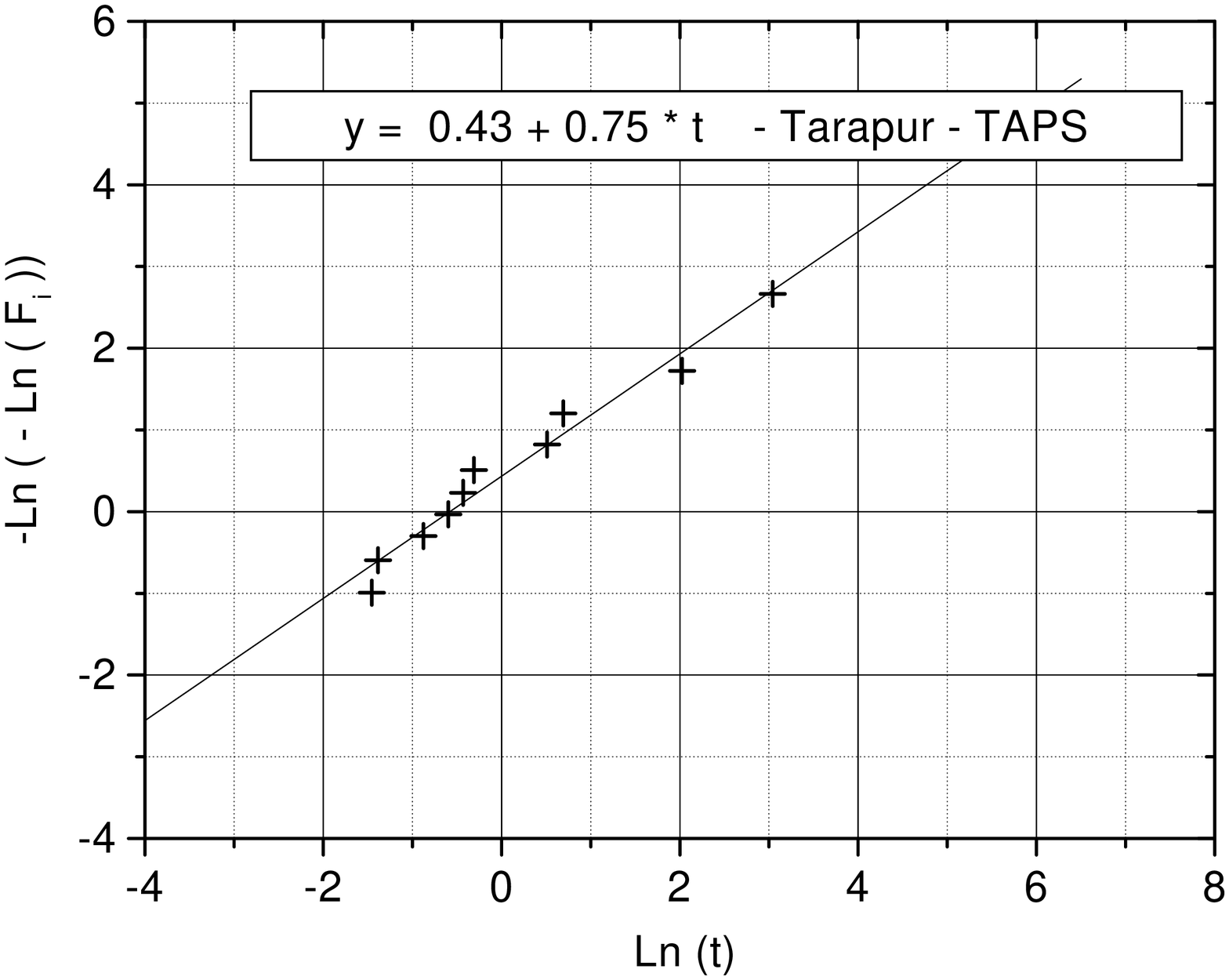}
\caption {\label{e} Log Log of $F_i$ versus Log of maximum annual
loss of off-site power duration for site Tarapur-TAPS}
\end{figure}
\begin{figure}
\includegraphics[width=3in,height=2in]{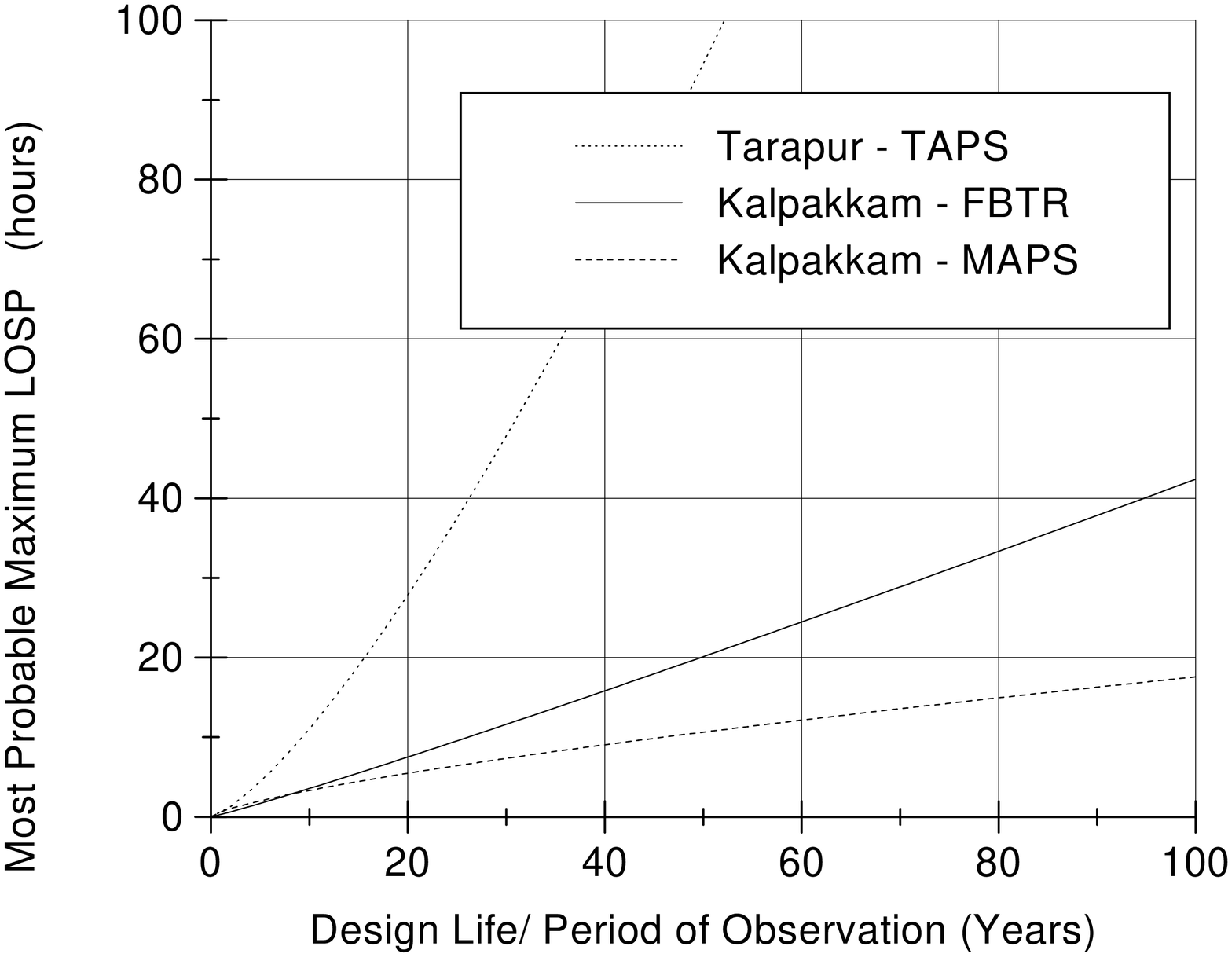}
\caption{\label{f} Most probable maximum LOSP duration versus time
of observation in years for the three nuclear plant locations}
\end{figure}


\end{document}